\documentstyle[aps,epsfig]{revtex}

\newcommand{\ket}[1]{%
| #1 \rangle 
}
\newcommand{\matrixel}[3]{%
\langle #1 | #2 | #3 \rangle 
}
\begin{document}


\title{New Criteria for Bosonic Behavior of Excitons}
\author{Monique Combescot}
\address{Groupe de Physique des Solides, Universit\'{e} Denis Diderot and Universit\'{e} Pierre et Marie Curie,\\
CNRS, Tour 23, 2 place Jussieu, 75251 Paris Cedex 05, France}
\author{Christian Tanguy}
\address{France Telecom R\&D RTA/CDP, 196 avenue Henri Ravera, 92225 Bagneux Cedex, France}
\maketitle
\begin{abstract}
Dimensional arguments lead to say that $N$ excitons in a sample of 
volume $V$ behave as bosons for $\lambda \, N a_x^3/V \ll 1$, $a_x$ 
being the exciton radius and $\lambda$ a dimensionless factor. The 
Mott criterion, which is based on the disappearance of all exciton 
bound states because of screening, gives $\lambda \simeq 1$. Since 
excitons feel each other through both Coulomb interaction and Pauli 
exclusion between their electrons and holes, criteria based on the 
underlying fermionic character of the exciton should be even more 
relevant. Two significant quantities are the boson departure $1-
[B,B^{\dagger}]$ and boson number $B^{\dagger} \, B$, where 
$B^{\dagger}$ is the exciton creation operator. Their expectation 
values in the $N$-exciton state are close to their boson values for 
$\lambda \sim 100$ and $\lambda \sim 50$, respectively. By contrast, 
the expectation value of $B^N \, (B^{\dagger})^N$ in the vacuum state 
is close to $N!$ for $25 \, N^2 \, a_x^3/V \ll 1$ only. This 
surprising $N^2$ dependence comes from the intrinsic $N$-body 
character of Pauli exclusion. Consequences of these new criteria on 
the Bose condensation of excitons are discussed.
\end{abstract}



When the density of electron-hole (e-h) pairs in semiconductors is very large, the carriers form an e-h plasma. Although the plasma's Coulomb energy is large, it remains small compared to its kinetic energy so that, in this regime, the Coulomb interaction can be treated as a perturbation (with possible high-order processes, such as RPA or ``bubble'' diagrams, in order to avoid spurious divergences\cite{Fetter}). The situation is quite different in the low-density regime, where one electron basically interacts with only one hole, giving birth to a hydrogenoid ``atom'' called exciton. Since bound states only appear if the Coulomb interaction between electron and hole is treated exactly, i.e., to all orders in perturbation, an exciton may be viewed as the result of a repeated interaction between electron and hole through ``ladder'' diagrams\cite{Mahan}.

From dimensional arguments, we expect the cross-over between the two regimes to appear when the distance between two excitons is of the order of their size, i.e., when $\lambda \, N a_x^3/V \simeq 1$, $a_x$ being the exciton Bohr radius, $N$ the number of excitons in the sample volume $V$, and $\lambda$ a dimensionless factor.

The high-density regime in which the Coulomb energy is dominated by the kinetic energy is valid when the parameter $r_s$, defined by $N \, \frac{4}{3} \, \pi (r_s \, a_x)^3/V = 1$, is small compared to 1. This leads to $\lambda = 4 \, \pi/3 \simeq 4$. The prefactor $\lambda$ can also be obtained by evaluating the carrier density for which, due to Coulomb screening, no excitonic bound state survives. The accepted result\cite{Klingshirn} for this Mott density leads to $\lambda \simeq 1$.

According to the spin-statistics theorem, excitons, which are composite particles made of two fermions, should behave as bosons. While the above $\lambda$'s have been deduced from considerations in which the Coulomb interaction plays a crucial role, criteria based on the underlying fermionic character of the excitons should be even more relevant. Indeed, excitons are not perfect bosons because they ``feel'' each other not only through Coulomb interaction but also through Pauli exclusion between their electrons and holes\cite{MCRC,Keldysh}.

In this Letter, we address the question: Up to what density can excitons be treated as bosons? We show that the criteria based on the fermionic character of excitons generate prefactors $\lambda$ larger than 1 by two orders of magnitude. This casts new light upon the possible observation of the challenging Bose condensation of excitons.

For the sake of simplicity, we will restrict here to excitons with zero total momentum. In terms of the free electron ($a^{\dagger}_{k}$) and hole ($b^{\dagger}_{-k}$) creation operators, the exact creation operator for the ground-state exciton reads
\begin{equation}
B^{\dagger} = \sum_k \, \phi_k \, a^{\dagger}_{k} \, b^{\dagger}_{-k},
\label{definition B croix}
\end{equation}
where $\phi_k$ is the ground-state exciton wave function in momentum space. From eq.~(\ref{definition B croix}), we get
\begin{equation}
D = 1-[B,B^{\dagger}] = \sum_k \, |\phi_k|^2 \, (a^{\dagger}_{k} \, a_{k} + b^{\dagger}_{-k} \, b_{-k}).
\label{definition D}
\end{equation}
Obviously, $D$ would vanish for perfect bosons. It is usually said\cite{Hanamura,Haug} that this operator, which measures the departure from bosonic behavior, ``is of the order of $N \, a_x^3/V$''. Such a claim pertains to an operator expectation value only, so that one must specify the state on which the expectation value is taken. For the $N$-free-pair state
\begin{equation}
\ket{\psi_{k_1 \, \ldots k_N}} = a^{\dagger}_{k_1} \, b^{\dagger}_{-k_1} \cdots a^{\dagger}_{k_N} \, b^{\dagger}_{-k_N}\ket{0}  , 
\label{Psi paires libres}
\end{equation}
we find
\begin{equation}
\matrixel{\psi_{k_1 \, \ldots k_N}}{D}{\psi_{k_1 \, \ldots k_N}} = 2 \, \sum_{i=1}^N  |\phi_{k_i}|^2 \simeq 2 \, N \, |\phi_{k=0}|^2
\label{Moyenne de D paires libres}
\end{equation}
if all $k_i$'s are close to zero. For $V$ sufficiently larger than $a_x^{3}$ --- so as to avoid boundary effects --- the bulk exciton ground state gives $|\phi_{k}|^2 = z/(1+k^2 \, a_x^2)^{4}$, where $z = 64 \, \pi \, a_x^3/V$. The expectation value of $D$ on $\ket{\psi_{k_1 \, \ldots k_N}}$ is thus less than 1 for $\lambda = 128 \pi \approx 400$.

Actually, the relevant state in this problem is not the $N$-free-pair state but the $N$-pair state constructed from the ground-state exciton
\begin{equation}
\ket{\Psi^{(N)}} = \frac{(B^{\dagger})^N \ket{0}}{\sqrt{\matrixel{0}{B^N  \, (B^{\dagger})^N }{0}}} .
\label{definition Psi N excitons}
\end{equation}
This state is the valid picture for the ground state of $N$ pairs in the low-density limit, as shown by Keldysh and Kozlov\cite{KK}. As the quantity
\begin{equation}
\matrixel{0}{B^N  \, (B^{\dagger})^N}{0} = N! \; F_N
\label{definition F_N}
\end{equation}
appears in any expectation value taken on $\ket{\Psi^{(N)}}$, we will start by its calculation.

If $B$ were a perfect boson operator, we would get $F_N \equiv 1$. From eq.~(\ref{definition B croix}) we find instead
\begin{equation}
\begin{tabular}{ccl}
$F_1$ & = & $\sigma_1$\\
$F_2$ & = & $\sigma_1^2-\sigma_2$\\
$F_3$ & = & $\sigma_1^3-3 \, \sigma_1 \, \sigma_2 + 2 \, \sigma_3$\\
$F_4$ & = & $\sigma_1^4-6 \, \sigma_1^2 \, \sigma_2 + 8 \sigma_1 \, \sigma_3 +3 \, \sigma_2^2 - 6 \, \sigma_4$\\
$F_5$ & = & $\sigma_1^5-10 \, \sigma_1^3 \, \sigma_2 +20 \, \sigma_1^2 \, \sigma_3+15 \, \sigma_1 \, \sigma_2^2$\\
 & & $-30 \, \sigma_1 \, \sigma_4-20 \, \sigma_2 \, \sigma_3 +24 \, \sigma_5$
\end{tabular}
\label{premiers F_N}
\end{equation}
and so on, in which we have set
\begin{equation}
\sigma_m = \sum_k \, |\phi_k|^{2 m} = z^{m-1} \, f(m).
\label{definition sigma_m}
\end{equation}
The normalization of $\phi_k$ implies $\sigma_1 = 1$. For the bulk exciton ground state, $f(m)= 16 \, (8 m-5)!!/(8 m-2)!!$. By carefully counting the number of occurences of cycles of length $2 \, m$ leading to $\sigma_m$, it is possible to show that the general expression of $F_N$ is
\begin{equation}
F_N = N! \, \sum_{[i_m]} \hskip3mm \prod_{m=1}^N \, \frac{1}{i_m!} \hskip3mm \bigg( \frac{(-1)^{m+1} \, \sigma_m}{m} \bigg)^{i_m} ,
\label{formule F_N}
\end{equation}
where the $[i_m]$ are determined by $\sum \, m \, i_m = N$. An even more compact expression of $F_N$ can be found\cite{note2}
\begin{eqnarray}
F_N & = & \frac{d^N \, \Xi}{d\eta^N} \bigg|_{\eta=0} , \label{F_N a partir de Xi}\\
\Xi(\eta) & = & \prod_{k} \, (1 + \eta \, |\phi_k|^{2}) . \label{formule Xi}
\end{eqnarray}
However, these expressions are of limited practical usefulness for $N \gg 1$: eq.~(\ref{formule F_N}) already contains more than $10^6$ different terms for $N=60$, and $\Xi$ is a complicated function of $\eta$. For numerical purposes, the recursion formula
\begin{equation}
F_N = \sum_{m=1}^{N} \, (-1)^{m+1} \, \frac{(N-1)!}{(N-m)!} \, \sigma_m \, F_{N-m}
\label{formule iterative F_N}
\end{equation}
with $F_0 \equiv 1$ turns out to be more convenient.

We first note that eq.~(\ref{definition F_N}) implies $F_N \geq 0$ while, from eqs.~(\ref{definition sigma_m})--(\ref{formule iterative F_N}), $F_N$ is a polynom of order $N-1$ in $z$. Since $\sigma_2 = 33 \, z/128 = (33 \, \pi/2) \, a_x^3/V$, we find that $F_2>0$ for $V^{1/3} = L > L_2 \simeq 3.7 \, a_x$. For large $N$, numerical calculations show that $F_N >0$ for $L > L_N \simeq 4 \, a_x \, N^{1/3}$. As expected, excitons should be a few Bohr radii apart to be considered as bulk excitons. For smaller volumes, excitons distort and the $\sigma_m$'s are modified accordingly.

From eqs.~(\ref{definition sigma_m})--(\ref{formule F_N}), we find that the $z$-expansion of $F_N$ starts as
\begin{equation}
F_N = 1-\sigma_2 \, N \, (N-1)/2 + {\rm O}\big( (N^2 \, z)^2 \big) ,
\label{F_N perturbatif}
\end{equation}
leading to a $N^2$ dependence of the correction to 1. The exact value of $\sigma_2$ given above shows that $F_N \simeq 1$ for $25 \, N^2 \, a_x^3/V \ll 1$. A simple way to understand that this $N^2$ dependence comes from the $N$-body character of Pauli exclusion is to return to the definition of $B^{\dagger}$. We have
\begin{eqnarray}
\matrixel{0}{B^N  \, (B^{\dagger})^N }{0} & = & \sum_{k_i,k'_i} \, \phi_{k_1}^* \, \phi_{k'_1} \cdots \, \phi_{k_N}^* \, \phi_{k'_N} \nonumber \\
 & & \times \, \matrixel{0}{a_{k_1} \, \cdots \, a_{k_N} \, a^{\dagger}_{k'_N} \, \cdots \, a^{\dagger}_{k'_1}}{0} \nonumber \\
 & & \times \, \matrixel{0}{b_{-k_1} \, \cdots \, b^{\dagger}_{-k'_1}}{0}.
\label{decomposition BN BcroixN}
\end{eqnarray}
The matrix elements in eq.~(\ref{decomposition BN BcroixN}) give $\pm 1$ (their product being +1) for \{$k_i$\}=\{$k'_i$\} and the $k_i$ all different; if not, they give 0. This allows to reexpress $F_N$ as
\begin{equation}
F_N = \sum_{k_1} \, |\phi_{k_1}|^2 \, 
\sum_{k_2 \neq k_1} \, |\phi_{k_2}|^2 \, 
\cdots \! \! 
\sum_{k_N \notin \{k_1, k_2, \ldots \, , k_{N-1}\}} \, |\phi_{k_N}|^2 .
\label{decomposition F_N simple}
\end{equation}
As expected from Pauli exclusion, eq.~(\ref{decomposition F_N simple}) merely states that the $N^{\rm th}$ exciton must be built with a wave vector different from those already used by previous excitons. Without this restriction, all the sums would be equal to 1, and so would $F_N$, as for perfect bosons.

The $N^2$ dependence of $F_N$ may be obtained easily from eq.~(\ref{decomposition F_N simple}) using a model calculation, which contains all the physics of the problem. An exciton is basically an object of spatial extension $a_x$, so that in a volume $V=L^3$ ($L \gg a_x$) the wave function in $k$-space has a width $1/a_x$, all the $k$'s being $2 \, \pi/L$ apart (see Fig.~\ref{figure 1}(b)). Instead of its exact value, let us take $\phi_k = \phi_0$ up to $k=1/a_x$, and 0 otherwise (see Fig.~\ref{figure 1}(c)). The normalization condition implies $N_0 \, \phi_0^2 =1$, where $N_0 = [(1/a_x)/(2 \, \pi/L)]^3$ is the number of $k$'s such that $\phi_k \neq 0$. For $N < N_0$ we get
\begin{eqnarray}
F_{N+1} & = & 1 \, (1-\phi_0^2) \, \cdots \, (1- N \, \phi_0^2) \label{F_N+1 a} \\
 & = & \phi_0^{2 N} \; \frac{\Gamma(\phi_0^{-2})}{\Gamma(\phi_0^{-2}-N)} . \label{F_N+1 b}
\end{eqnarray}
For $N^2 \, \phi_0^2 \ll 1$, eq.~(\ref{F_N+1 a}) gives $F_N \simeq 1- N \, (N-1) \, \phi_0^2/2$ in agreement with eq.~(\ref{F_N perturbatif}), whereas eq.~(\ref{F_N+1 b}) gives 
\begin{equation}
F_N \simeq e^{- N \, (N-1) \, \phi_0^2/2} \hskip10mm {\rm for \ }N \, \phi_0^2 \ll 1.
\label{F_N exponentiel}
\end{equation}
Equation~(\ref{F_N exponentiel}) shows that although the excitons are ``close'' to bulk excitons for $N \, \phi_0^2 \ll 1$, their normalization factor $\matrixel{0}{B^N  \, (B^{\dagger})^N }{0}$ is much smaller than $N!$ (as for perfect bosons) if $N \, \phi_0^2 \ll 1 \ll N^2 \, \phi_0^2$. This $N^2$ dependence may be viewed as a {\em ``stimulated'' many-body effect}. Indeed, Pauli exclusion being intrinsically a N-body interaction, many-body effects based on this interaction are quadratic. Note that this exponentially small factor $F_N$ turns out to have a relatively small influence on $N!$ because
\begin{equation}
N! \, F_N \simeq e^{N \, (\ln N -1 -(N-1) \, \phi_0^2/2+ \, \cdots)} .
\label{F_N et Stirling}
\end{equation}
If we now consider the exact $\phi_k$, we can actually show that $F_N$ has a similar behavior
\begin{equation}
F_N \simeq e^{- N \, (N-1) \, \sigma_2/2} \hskip10mm {\rm for \ }N \, \sigma_2 \ll 1 ,
\label{F_N exponentiel general}
\end{equation}
in agreement with the first terms of the expansion given in eq.~(\ref{F_N perturbatif}).

We now turn to the expectation value of $D$ taken on $\ket{\Psi^{(N)}}$. It is convenient to introduce $C^{\dagger}$ defined by $[D,B^{\dagger}] = 2 \, C^{\dagger}$. Since $[C^{\dagger}, B^{\dagger}]=0$ and $D \ket{0} = 0$, we get
\begin{equation}
D \, (B^{\dagger})^N \, \ket{0} = 2 \, N \, (B^{\dagger})^{N-1} \, C^{\dagger} \, \ket{0} .
\label{D B croix N}
\end{equation}
This leads to
\begin{eqnarray}
B \, (B^{\dagger})^{N+1} \, \ket{0} & = & (N+1) \, (B^{\dagger})^{N} \, \ket{0} \nonumber \\
 & & - N \, (N+1) \, (B^{\dagger})^{N-1} \, C^{\dagger} \, \ket{0} .
\label{B B croix N}
\end{eqnarray}
From eqs.~(\ref{D B croix N}) and (\ref{B B croix N}), we easily deduce
\begin{equation}
\langle D \rangle_{{}_N} = \frac{\matrixel{0}{B^N  \, D \, (B^{\dagger})^N }{0}}{\matrixel{0}{B^N  \, (B^{\dagger})^N }{0}} = 2 \, \bigg( 1-\frac{F_{N+1}}{F_{N}} \bigg) .
\label{moyenne D}
\end{equation}

We check that $\langle D \rangle_{{}_N}=0$ for perfect bosons. For our model $\phi_k$, eq.~(\ref{F_N+1 a})
immediately gives $\langle D \rangle_{{}_N}= 2 \, N \, \phi_0^2$, while for the exact exciton wave function, we find
\begin{equation}
\langle D \rangle_{{}_N} \simeq 2 \, N \, \sigma_2 \hskip10mm {\rm for \ }N \, \sigma_2 \ll 1 .
\label{moyenne D vrai exciton}
\end{equation}
Taking $\langle D \rangle_{{}_N} \ll 1$ as the criterion for $N$ bulk excitons to behave as bosons leads therefore to a prefactor $\lambda = 33 \, \pi \simeq 100$.

Another relevant operator is the boson number in the $\ket{\Psi^{(N)}}$ state. From eqs.~(\ref{definition B croix}) and (\ref{moyenne D}), it is straightforward to show that
\begin{equation}
\langle B^{\dagger} \, B \rangle_{{}_N} = N - (N-1) \, \langle D \rangle_{{}_N}/2 .
\label{moyenne Bcroix B}
\end{equation}
Obviously, $\langle B^{\dagger} \, B \rangle_{{}_N} = N$ for perfect bosons because $\langle D \rangle_{{}_N} \equiv 0$. Its value is $N- N\, (N-1) \, \phi_0^2$ with our model $\phi_k$ while from eq.~(\ref{moyenne D vrai exciton}) we get
\begin{equation}
\langle B^{\dagger} \, B \rangle_{{}_N} \simeq N \, (1 - (N-1) \, \sigma_2) \hskip6mm {\rm for \ }N \, \sigma_2 \ll 1 .
\label{moyenne Bcroix B vrai exciton}
\end{equation}
The percentage of pairs not counted as bosons is thus small for $N \, \sigma_2 \simeq 50 \, N \, a_x^3/V \ll 1$.

The above formalism can be readily applied to 2D excitons. From the 2D ground state wave function we have $|\phi_k|^{2} = y/(1+k^2 \, a_x^2/4)^3$, where $y=2 \, \pi \, a_x^2/S$, $S$ being the sample's area and $a_x$ still the 3D exciton Bohr radius. We find $\sigma_m = 2 \, y/(3 \, m -1)$, so that $\sigma_2 = 2 \, y/5$. Numerical calculations give $L_N \simeq 1.5 \, a_x \, N^{1/2}$ for $N \gg 1$, while the condition $2 \, N \, \sigma_2 \ll 1$ now reads $5 \, N \, a_x^2/S \ll 1$. Assigning to the 2D exciton an effective radius $a_{\rm 2D}=a_x/4$, which corresponds to the maximum of the radial charge density\cite{Chemla}, we find again that our new criterion differs from the geometrical one $N \, a_{\rm 2D}^2/S \ll 1$ by nearly two orders of magnitude.

Let us end by reconsidering the possibility of Bose-Einstein condensation of 3D excitons in the light of these new criteria. $N$ noninteracting bosons undergo Bose-Einstein condensation when their density $n=N/V$ exceeds a temperature-dependent critical value $n_c(T)$, which, when applied to excitons, is given by \cite{Keldysh}
\begin{equation}
\frac{k_B \, T}{R} = 6.62 \, \frac{m_e \, m_h}{(m_e + m_h)^2} \, \left(\frac{a_x^3 \, n_c(T)}{g}\right)^{2/3},
\label{T critique}
\end{equation}
where $R$ and $g$ are the exciton binding energy and degeneracy, respectively. For Cu${}_2$O orthoexcitons\cite{Goto}, $R \simeq 150 {\rm\ meV}$, $a_x \simeq 0.7 {\rm\ nm}$, $g=3$, and $m_e \sim 1$, $m_h \sim 0.6$, so that $100 \, n_c \, a_x^3 \simeq 0.1$ for a carrier temperature $T = 10 {\rm K}$, which means that for $n>n_c$, excitons barely obey our criterion $100 \, n \, a_x^3 \ll 1$ for bosonic behavior. Besides, considering a typical excitation volume $V \sim (10 \, \mu{\rm m})^2 \times (1 \, \mu{\rm m})$ necessary to get a large enough density, we find $25 \, n_c^2 \, V \, a_x^3 \sim 10^{7}$. On account of our criterion $25 \, N^2 \, a_x^3/V \ll 1$, $F_N$ would be extremely small. It thus seems unlikely that excitons could be considered as noninteracting bosons under these experimental conditions. The critical density should therefore differ from eq.~(\ref{T critique}). In contrast, let us stress that both our criteria are fulfilled in the case of atoms when we put numbers corresponding to actually observed Bose-Einstein condensation. Indeed, for $a_0 \sim 0.1 \, {\rm nm}$ and condensates of up to $5 \times 10^6$ atoms, we find at most $100 \, N \, a_0^3/V \sim 10^{-8}$ and $25 \, N^2 a_0^3/V \sim 10^{-3}$ for the different configurations reported in the literature\cite{BEC atomique}. The main reason for such a difference lies with the temperature $T$, which is about $10^{-7} \, {\rm K}$ in atomic condensates. Excitons being ``transient'' particles produced by excitations, they cannot be cooled down as much.

\vskip5mm
\noindent {\bf Conclusion}
\vskip0.3cm

While previous criteria for bosonic behavior of excitons were related to the Coulomb interaction between electron-hole pairs, we have studied the consequences of the fermionic character of these excitons induced by Pauli exclusion between their constituents. They lead to prefactors two orders of magnitude larger than the usual Mott criterion $N \, a_x^3/V \simeq 1$, so that excitons can be considered as bosons up to densities well below the Mott density only. This is of importance for the possible observation of the Bose condensation of excitons. We have also found that $\matrixel{0}{B^N \, (B^{\dagger})^N }{0}$ is exponentially smaller than its boson value $N!$  unless $25 \, N^2 \, a_x^3/V \ll 1$. This unexpected $N^2$ dependence has to be related to the intrinsic $N$-body character of Pauli exclusion.

\vskip5mm
\noindent {\bf Acknowledgements}

We gratefully acknowledge very helpful discussions with Bernard Roulet, Roland Combescot, Claude Benoit \`{a} la Guillaume, Philippe Lavallard, and Catherine Gourdon.

\begin{figure}
\caption{
Representation of the ground-state excitonic wave function in real space (a) and $k$-space (b). The model wave function is represented in (c).
}
\label{figure 1}
\end{figure}

\vfill

\hfil\includegraphics{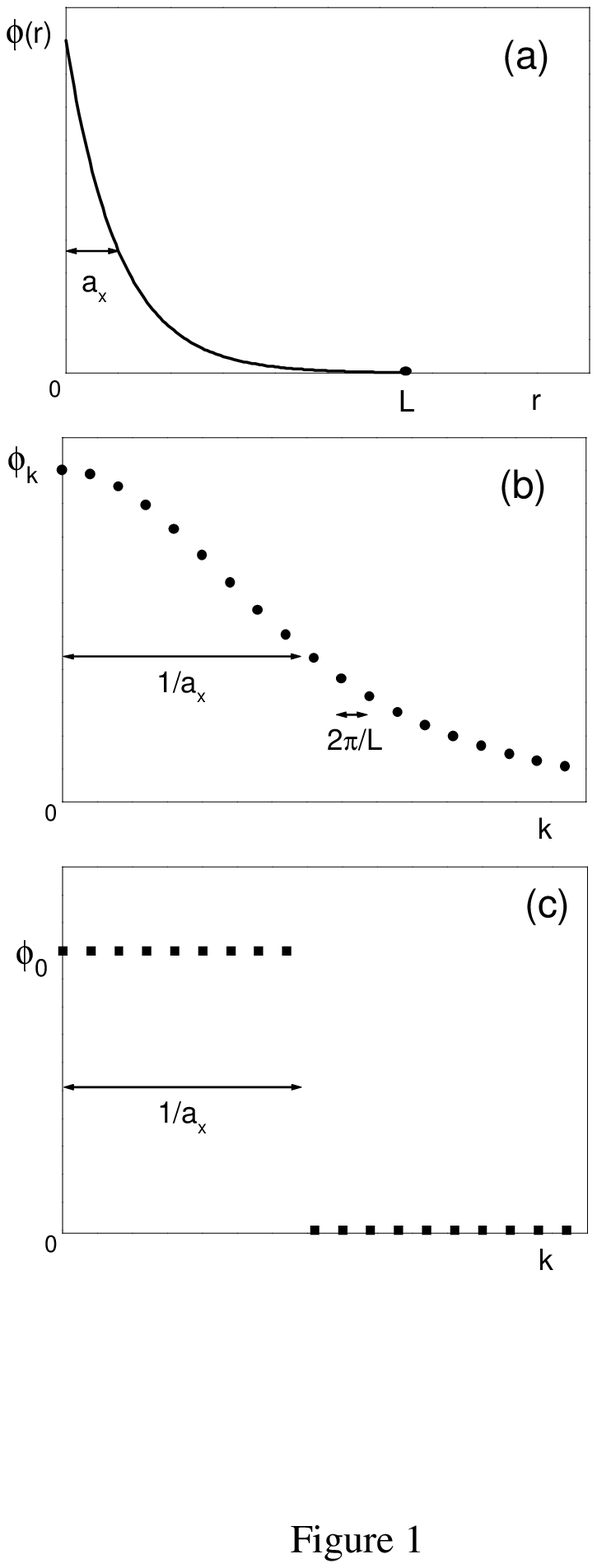}\hfil


\begin{references}
\bibitem{note2} If the number of ``sites'' $k$ is less than $N$, $F_N=0$, as expected for fermions.
\end{references}
\end{document}